\documentclass[10pt]{article}
\usepackage[letterpaper]{geometry}
\usepackage{hicss}
\usepackage{times}
\usepackage[none]{hyphenat}
\usepackage{url}
\usepackage{latexsym}
\usepackage{xcolor}
\usepackage{xspace}
\usepackage{indentfirst}
\usepackage{graphicx}
\graphicspath{{images/}}
\usepackage[
    style=apa,
  ]{biblatex}
\usepackage[toc, acronym, nopostdot,nonumberlist]{glossaries}
\newacronym{der}{DER}{Distributed Energy Resource}
\newacronym{ders}{DERs}{Distributed Energy Resources}
\newacronym{pel}{PEL}{power-electronics}

\title{CAMEO: A Co-design Architecture for Multi-objective Energy System Optimization}
\author{Rounak Meyur, Tonya Martin and Sumit Purohit \\
Pacific Northwest National Laboratory\\
{\{rounak.meyur,tonya.martin,sumit.purohit\}@pnnl.gov}}

\date{June 2024}

\newcommand{\cameo}{\textsc{CAMEO}\xspace}
\newcommand{\nextflow}{Nextflow\xspace}
\newcommand{\docker}{{docker}\xspace}

\newcommand{\wind}{\textsc{wind}\xspace}
\newcommand{\battery}{\textsc{battery}\xspace}
\newcommand{\sset}{\textsc{scen\_set}\xspace}
\newcommand{\stree}{\textsc{scen\_tree}\xspace}
\newcommand{\dsset}{\textsc{design\_ss}\xspace}
\newcommand{\dstree}{\textsc{design\_st}\xspace}
\newcommand{\summary}{\textsc{summarize}\xspace}

\begin{document}

\maketitle

\begin{abstract}
Co-design plays a pivotal role in energy system planning as it allows for the holistic optimization of interconnected components, fostering efficiency, resilience, and sustainability by addressing complex interdependencies and trade-offs within the system. 
This leads to reduced operational costs and improved financial performance through optimized system design, resource allocation, and system-wide synergies. 
In addition, system planners must consider multiple probable scenarios to plan for potential variations in operating conditions, uncertainties, and future demands, ensuring robust and adaptable solutions that can effectively address the needs and challenges of various systems. 
This research introduces Co-design Architecture for Multi-objective Energy System Optimization (\cameo), which facilitates design space exploration of the co-design problem via a modular and automated workflow system, enhancing flexibility and accelerating the design and validation cycles. The cloud-scale automation provides a user-friendly interface and enable energy system modelers to efficiently explore diverse design alternatives. \cameo aims to revolutionize energy system optimization by developing next-generation design assistant with improved scalability, usability, and automation, thereby enabling the development of optimized energy systems with greater ease and speed. 
\end{abstract}

\subsubsection*{Keywords: co-design optimization, co-design architecture, modular co-design workflow, scalable co-design framework, electric storage planning, \docker, \nextflow}

\maketitle

\section{Introduction}

\gls{ders} are becoming increasingly integral to the modern power grid, transforming the landscape of energy generation, transmission, and distribution. 
Renewable energy sources, particularly wind and solar, are at the forefront of this transformation~(\cite{kataray2023}). 
However, the intermittent nature of these resources presents significant challenges. 
To address the variability in power generation, energy storage systems, specifically batteries, are essential~(\cite{degue2023}). 
These storage systems not only stabilize the supply but also enable participation in energy markets, thereby adding an economic dimension to their utility~(\cite{heredia2015}).

The financial viability of integrating battery storage into the power grid is based on determining the optimal size of the battery, which involves substantial initial investment(~\cite{kitner2012}). 
Effective participation in various energy markets, such as the day-ahead, real-time, and reserve markets, depends on this optimization. 
A co-design approach to optimization is crucial, as it simultaneously considers multiple objectives to maximize revenue and ensure efficient market participation~(\cite{heredia2018}).
Current optimization capabilities~(\cite{yin2023,jiaxin2023}) are often specialized for particular scenarios, which restrict scalability, heterogeneity, and usability, especially in high-penetration \gls{pel} energy systems. 



This paper presents \cameo, a co-design architecture for multi-objective energy system optimization. 
The proposed architecture tackles key challenges in \emph{scalability}, enabling the system to manage increased complexity without proportionate computational resource expenditure. 
It also ensures \emph{heterogeneity} by allowing diverse components and objectives within the same framework.
In addition, user-friendly interfaces and standardized workflows are developed to enhance \emph{usability}.
We also present a use case to design offshore wind farms, with the objective of balancing the economic and operational aspects of integrating \gls{ders} and energy storage into the modern power grid.

\noindent\textbf{Contributions.}~ 
\cameo is a framework to aid researchers and system planners in running an optimization formulations developed a priory, over a large design parameter space.
The contributions of \cameo can be summarized as follows:
(i) \textbf{Scalable framework for design-space exploration}: For a given optimization formulation, \cameo explores a wide design (hyper)-parameter space in high performance computing environment. \cameo identifies the combinations of input parameter configurations and performs parallel execution of the given optimization problem using the available computing resources. 
(ii) \textbf{Modular approach for enhanced heterogeneity}: \cameo employs a modular approach to facilitate multiple instances of the workflow, allowing it to be easily adapted for several optimization formulations at run-time for a given co-design problem.
This flexibility enables users to tailor the framework to specific needs and objectives, promoting broad applicability across different energy system optimization scenarios.
(iii) \textbf{Containerized architecture for enhanced usability}: \cameo provides an automated workflow of optimization formulations in dedicated \docker containers, offering a lightweight and portable solution for deployment and execution. 
A containerized architecture ensures consistent performance across different environments, simplifies dependency management, and reduces the overhead associated with traditional virtual machines. 
This approach enables quick setup, scalable operations, and efficient resource utilization, making \cameo highly accessible and user-friendly for diverse energy system co-design applications.

The remainder of the paper is outlined as follows: Section~\ref{sec:rel} gives an overview of similar frameworks and their drawbacks. 
Thereafter, Section~\ref{sec:arch} focuses on the architecture of the proposed \cameo framework and address key research questions regarding its design.
We illustrate an energy system co-design optimization use case of \cameo in Section~\ref{sec:use}.

\section{Related Works}\label{sec:rel}
Co-design has been extensively used in different science and technology domains. 
In contrast to a sequential approach of designing multiple optimal design parameters, co-design provides a system-wide optimal guarantee including design and control \cite{fathy2001coupling}. 
Particularly, \gls{pel}-enabled energy systems present novel characterization, computational, and optimization challenges in the presence of multiple conflicting objectives. 
Current research landscape lacks a standardized approach to characterizing multi-objective, multi-scale optimization problems.    

\cite{guo2014metronomy} have developed `Metronomy', a co-simulation driven co-design architecture to perform design space exploration. 
`Metronomy' focuses on timing properties such as latency from sensing to actuation on the control performance in an electrical power system use case.
\cite{zheng2016convince} have proposed a design exploration framework for connected vehicles called `CONVINCE'. 
The framework includes different mathematical models, simulators, and validation algorithms to model inter-vehicle communications and software and hardware design parameters. 
\cite{ramachandran2023computational}  present a simulation-based control co-design computational framework that designs physical components and control parameters of a microgrid system for resilience objectives. 

In contrast to these use-case oriented solutions, \cameo identifies common design patterns and components of a multi-objective co-design problem and develop a modular plug-and-play and generalized solution to assist system planners.
Our proposed approach provides a robust modular workflow architecture and intuitive user interfaces, which collectively support dynamic and customizable co-design workflows across diverse optimization paradigms and data environments.

\section{Architecture and Methodology}\label{sec:arch}
\begin{figure*}[tbhp]
    \centering
    \includegraphics[width=0.99\textwidth]{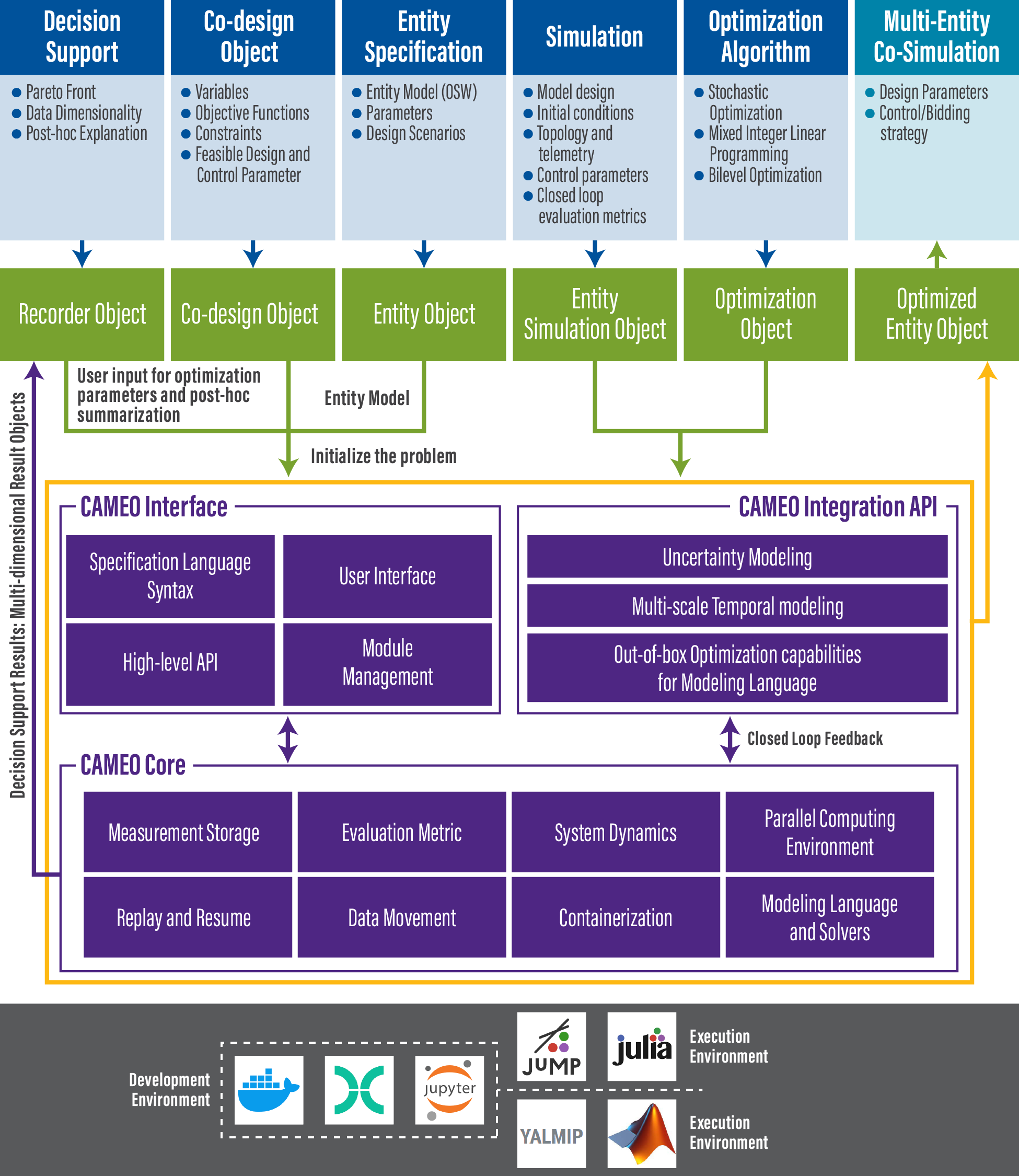}
    \caption{\cameo Architecture: Modular Design to support multiple instances of co-design workflows}
    \label{fig:cameo-arch}
\end{figure*}
\cameo is a use-case agnostic co-design framework that provides scalable and modular ways to explore the design-space of an optimization formulation. The framework executes multiple instances of the formulation in parallel to provide insight into the accuracy, sensitivity, and computing performance of the formulation in a high performance computing environment.  

As shown in Figure~\ref{fig:cameo-arch}, \cameo uses a modular approach to define various components and their behavior. 
For example, an entity object is used to specify design parameters and hyper-parameters. 
Similarly, the optimization object links the entity object to an algorithm and corresponding solver with seed parameters. 
Furthermore, a simulation object defines the underlying energy system topology, telemetry, control parameters, and evaluation metrics. \cameo also monitors the execution environment and generates provenance summaries to share with downstream decision support applications such as pareto-front visualizations and dashboards. 
An instance of the \cameo architecture is implemented as a cloud-scale workflow system using a subset of the objects available as part of the architecture. 
Presently, the \cameo workflow framework is available for \textsc{Python} and \textsc{Julia} environment. 
In the future, \textsc{Matlab} will be added as a target execution environment.

\cameo addresses following key research questions pertaining to a co-design problem formulation:
\begin{itemize}
    \item How do we characterize the different ways in which a co-design problem can be specified and solved?
    \item What is the minimum descriptive language required to define a single entity, multi-objective, multi-scale co-design problem.
    \item What is the required computational architecture and challenges to implement a modular and scalable co-design workflow system.
\end{itemize}

\subsection{Characterizing approaches}
For a single entity co-design, \cameo defines a \emph{quasi-model} to identify all the components of the co-design process. 
Thereafter, it becomes easier to systematically integrate various elements such as models, control parameters, and design data. 
This ensures that all necessary aspects of the co-design problem are considered and appropriately linked.

It supports multi-modal, multi-temporal scale \emph{data flow} between components such as model, control parameters, design and operational data, constraints, objective functions, dependencies, sub-system topologies, parameters, and hyper parameters as shown in Figure~\ref{fig:cameo-arch}.
This allows for the seamless exchange of information and ensures that data from different sources and of different types can be incorporated into the workflow, enhancing the robustness and accuracy of the co-design process.

The \emph{modular plug-and-play design} enables dynamic definition of multiple co-design workflow instances by combining different models, objectives, optimization formulations, and co-design approaches. 
This provides the versatility needed to tackle various co-design problems effectively.

\emph{Standardized interfaces} and APIs across components ensure extensive compatibility with various data formats and validation schema.
This standardization simplifies the integration process, and ensures that different components can work together seamlessly.

Additionally, \cameo supports \emph{multiple optimization paradigms}, including white-box and simulation-based optimization, various programming environments, and AI/ML frameworks.
This adaptability makes it suitable for a wide range of co-design problems and the framework can be tailored to specific needs and preferences.
It also provides \emph{simple user interfaces} for workflow definitions, and visualizations for time series data, hierarchical subsystems, and multi-layer attributed graphs.
This makes the co-design process more accessible, where users can easily define, monitor and adjust workflows, leading to efficient problem solving.
This characterization promotes a structured yet flexible approach to creating modular workflow processes, making it easier to address the complex and heterogeneous nature of co-design problems.

\subsection{Minimum descriptive language}
\cameo provides a declarative, multi-domain modeling language to specify the co-design problem in a machine-readable form.
This language is orchestrated via a \emph{cloud scale workflow syntax}, allowing for scalable and efficient processing by leveraging cloud resources for large-scale optimization tasks.

It emphasizes developing an \emph{interoperable integration specification}, allowing different models and optimization approaches to seamlessly share inputs, outputs, and dependencies. 
This modularity ensures that different components can be easily integrated and interchanged, making the workflow adaptable to diverse co-design problems.
Validating a co-design problem based on selected parameters and optimization approaches for entity and system models under given constraints is crucial. 
\cameo exports component specifications as contracts and validates the workflows based on these contracts, addressing interoperability challenges by describing inter-subsystem dynamics and dependencies. 
This reduces errors and enhances the robustness of the workflow, making it more user-friendly.

Additionally, it uses a flexible \textsc{json}-based hierarchical serialization format for encoding and persisting various design variables, making it easier to manage and modify input data for the simulation-based models.
This enhances usability by providing a straightforward and widely understood data format.

Another major aspect of \cameo is that it serves as a tool that assists users in selecting from various use cases, entities, objectives, constraints, and initial parameters to solve domain-specific co-design problems. 
Specifically tailored for applications like the wind farm design use case described in section \ref{sec:use}, the framework targets users such as researchers, consulting power electronics engineers, and policy makers proficient in defining co-design problems and analyzing the optimal solutions.
In the future, we will also develop a workflow-driven user interface (UI) for the user selections and will produce optimal solutions encoded as multi-objective solution set such as pareto front.

\begin{table*}[tbhp]
\centering
\caption{Comparison of different scientific workflow systems.}
\label{tab:workflow}
\begin{tabular}{|l|l|l|l|l|l|}
\hline
 &
  \textbf{Language} &
  \textbf{Parallelization} &
  \textbf{Flow Control} &
  \textbf{\begin{tabular}[c]{@{}l@{}}Containers \\ Supported\end{tabular}} &
  \textbf{\begin{tabular}[c]{@{}l@{}}Cloud \\ Platforms\end{tabular}} \\ \hline
\textbf{Nextflow} &
  \begin{tabular}[c]{@{}l@{}}\textsc{groovy},\\ \textsc{java}\end{tabular} &
  \begin{tabular}[c]{@{}l@{}}Configurable with \\ automatic retries\end{tabular} &
  \begin{tabular}[c]{@{}l@{}}Workflow definition files \\ and variables, command \\ line parameter settings.\end{tabular} &
  \begin{tabular}[c]{@{}l@{}}\docker, \\podman,\\ \textsc{singularity}\end{tabular} &
  \begin{tabular}[c]{@{}l@{}}AWS, \\Azure, \\ Google Cloud, \\ Kubernetes\end{tabular} \\ \hline
\textbf{Snakemake} &
  \textsc{python} &
  \begin{tabular}[c]{@{}l@{}}Configurable with \\ specific retires\end{tabular} &
  \begin{tabular}[c]{@{}l@{}}File inputs and outputs, \\ command line CPU settings\end{tabular} &
  \begin{tabular}[c]{@{}l@{}}\docker \\through\\ \textsc{singularity}\end{tabular} &
  Kubernetes \\ \hline
\textbf{\textsc{airflow}} &
  \textsc{python} &
  \begin{tabular}[c]{@{}l@{}}Configurable at \\ Scheduler, or at \\ task level\end{tabular} &
  \begin{tabular}[c]{@{}l@{}}Directed Acyclic Graph \\ (DAG) based pipeline \\ defined in \textsc{python} script\end{tabular} &
  \begin{tabular}[c]{@{}l@{}}\docker, \\ \textsc{singularity}\end{tabular} &
  \begin{tabular}[c]{@{}l@{}}AWS, \\Azure,\\ Google Cloud,\\ Kubernetes\end{tabular} \\ \hline
\end{tabular}
\end{table*}

\subsection{Co-design workflow system}
We use \nextflow~(\cite{di2017nextflow}), a scientific workflow system to demonstrate \cameo implementation. 
The framework is instantiated as a containerized and configurable execution platform with relevant technology stack, standardize interfaces, optimal data formats, and validation schema.
\nextflow provides portability and reproducibility through wide support for container technologies on multiple executors for \textsc{SLURM}, \textsc{Moab} and batch schedulers, as well as for \textsc{Kubernetes} and cloud platforms.  
\nextflow pipelines integrate with most scripting languages to allow existing code reuse resulting in fast prototyping.  
Intermediate results are tracked and checkpointed allowing processes to resume for the last successful step and provide process provenance.

A comparison of \nextflow with its close competitions `Snakemake' and `Apache \textsc{airflow}' is shown in Table~\ref{tab:workflow}.
We prefer \nextflow due to its comprehensive support for multiple container tools and flexible execution across various environments and cloud platforms. 
Its command line parameter settings and versatile definition files allow seamless transition from development to production workflows within the same framework. 
Moreover, \nextflow provides robust provenance tracking and detailed execution reporting, enhancing transparency and reproducibility across workflow executions. 
These features make Nextflow particularly advantageous for managing complex workflows with diverse computational requirements and deployment scenarios.

\section{Use Case Demonstration} \label{sec:use}
In this section, we illustrate a power grid planning use case, where we utilize \cameo to perform a co-design over an extensive parameter space.
We briefly explain the problem, followed by describing the various building blocks used to construct the workflow.
Finally, we show the results obtained from two optimization formulations and various statistics of their corresponding computational overhead.

\subsection{Co-design optimization problem}
\begin{figure}[tbhp]
    \centering
    \includegraphics[width=0.48\textwidth]{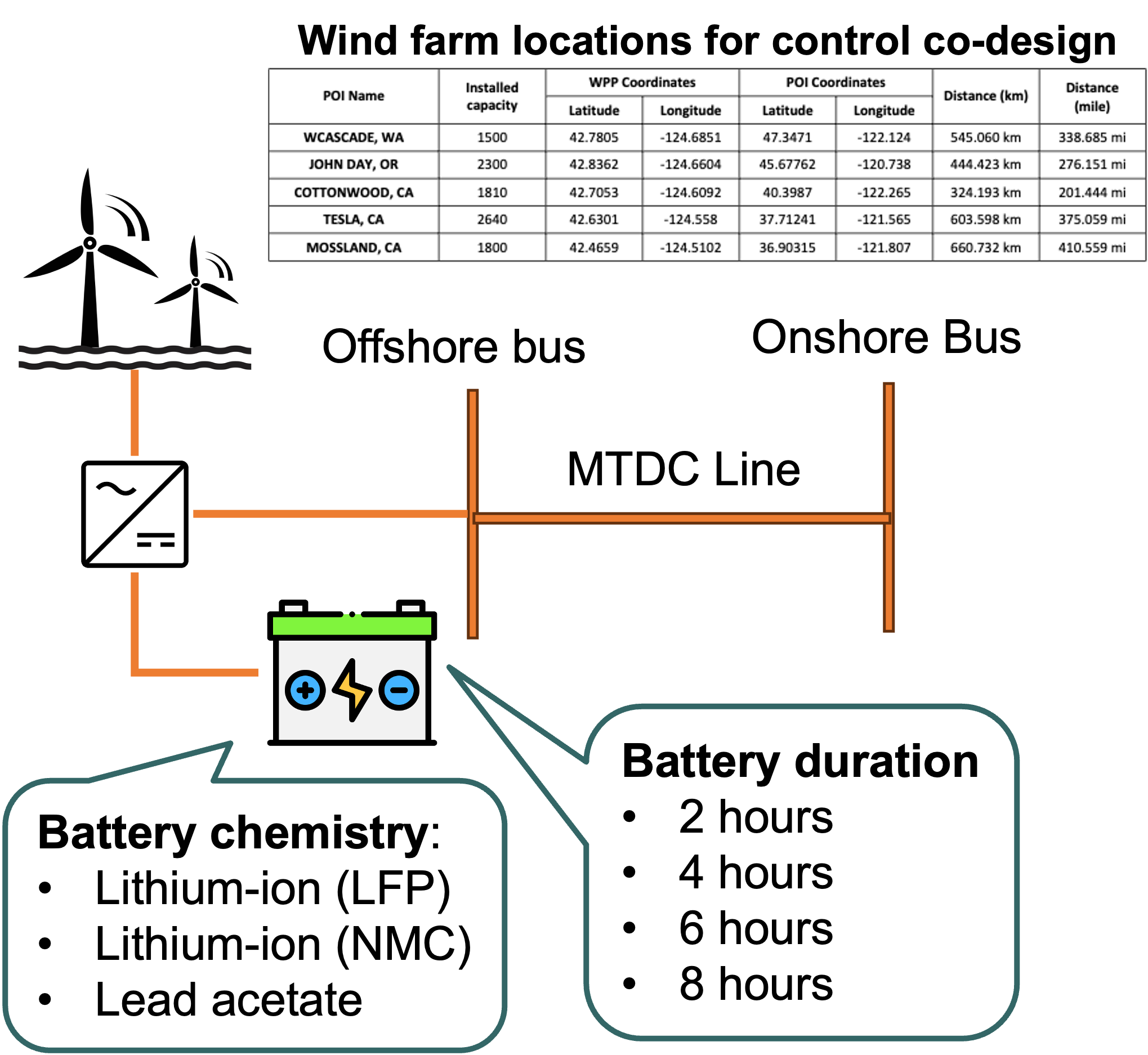}
    \caption{\cameo use case: optimal energy storage sizing problem needs to be executed for the various battery configurations and wind farm locations.}
    \label{fig:cameo-prob}
\end{figure}
The objective of the co-design problem is to optimize the energy storage design for wind farms. 
The goal is to maximize revenue returns from day-ahead, real-time, and reserve energy markets, with an initial installation investment. 
The optimization problem involves determining the optimal battery size based on the power grid topology, historical energy prices, and wind speed data. 
Despite the fixed problem formulation, \cameo is capable of handling multiple input parameter configurations, including variations in the location of wind farm, the chemistry of the battery, and the duration of the battery(as shown in Figure~\ref{fig:cameo-prob}). 
These variations influence the resulting optimal battery size. 
Further, we have two stochastic optimization formulations which solve this co-design problem~(\cite{himanshu2024}) -- (i) \textbf{Formulation A} which is based on generating multiple sets of random scenarios and maximizing the expected revenue objective for each random set, and (ii) \textbf{Formulation B} which is based on constructing \emph{scenario trees}~(\cite{Kirui2020}) and solving a multi-stage stochastic optimization problem.

\cameo excels in systematically executing the optimization across the diverse parameter sets, generating detailed summaries of the optimization results and providing information about the computational overhead for each formulation.
This capability ensures scalability and efficiency in evaluating numerous scenarios, providing valuable insights for strategic decision-making in energy storage planning for the different wind farm location.

\subsection{\nextflow process definitions}
In this section, we describe the various processes defined using \nextflow. 
These processes form the building blocks for the entire co-design optimization workflow.
Each of the processes are identified by a name which is denoted within squared braces.
For the remainder of this section, we will use these names as placeholders for the corresponding \nextflow process.

\noindent\textbf{Wind farm data collection [\textsc{wind}].}~
The process gathers comprehensive wind farm data, extracting crucial information such as the geographical coordinates (longitude and latitude) of the various wind farm locations. 
Additionally, it collects historical wind speed data~(\cite{nrel_wind}) and energy prices at the points of interconnection~(\cite{caiso}), providing the required dataset for further analysis and optimization.

\noindent\textbf{Battery configuration collection [\textsc{battery}].}~
The process gathers detailed battery configuration data, extracting information such as battery chemistry, power rating, and battery duration. 
Additionally, it collects data on the installation cost in dollars per kilowatt (\$/kW) and the battery efficiency, also known as round trip efficiency (RTE).

\noindent\textbf{Random scenario set generation [\textsc{scen\_set}].}~
The process generates random scenario sets, each comprising wind speed and energy price data for multiple representative days. 
Utilizing the collected historical data for a specific wind farm location, it creates multiple scenario sets, with the exact number determined by user input.

\noindent\textbf{Scenario tree generation [\textsc{scen\_tree}].}~
The process generates a scenario tree for a multi-stage stochastic optimization problem, leveraging the collected historical data for a specific wind farm location. 
This tree structure facilitates the modeling of uncertainties and decision-making stages over multiple time scales (for the different energy market time scales).

\noindent\textbf{Scenario set based design [\textsc{design\_ss}].}~
The process solves a constrained revenue maximization problem for each generated random scenario set. 
Each problem incorporates standard battery and power grid constraints, ensuring feasible and efficient solutions tailored to the scenario set.
This process forms the core component of Formulation A.

\noindent\textbf{Scenario tree based design [\textsc{design\_st}].}~
The process solves a multi-stage optimization problem for each generated scenario tree, focusing on constrained revenue maximization. 
Each problem adheres to the battery and power grid constraints, ensuring that the solutions are both practical and optimized for long-term performance and profitability.
This process forms the core component of Formulation B.

\noindent\textbf{Summarizing and visualization [\textsc{summarize}].}~
The process summarizes the results of the optimization design by compiling them into a consolidated CSV file and generating visual plots for clear interpretation. 
This step enables stakeholders to comprehensively review and analyze the outcomes of the energy system co-design optimization, facilitating informed decision-making and further refinement of the integrated energy solutions.

\subsection{Results from Formulation A.}
\begin{figure}[tbhp]
    \centering
    \includegraphics[width=0.48\textwidth]{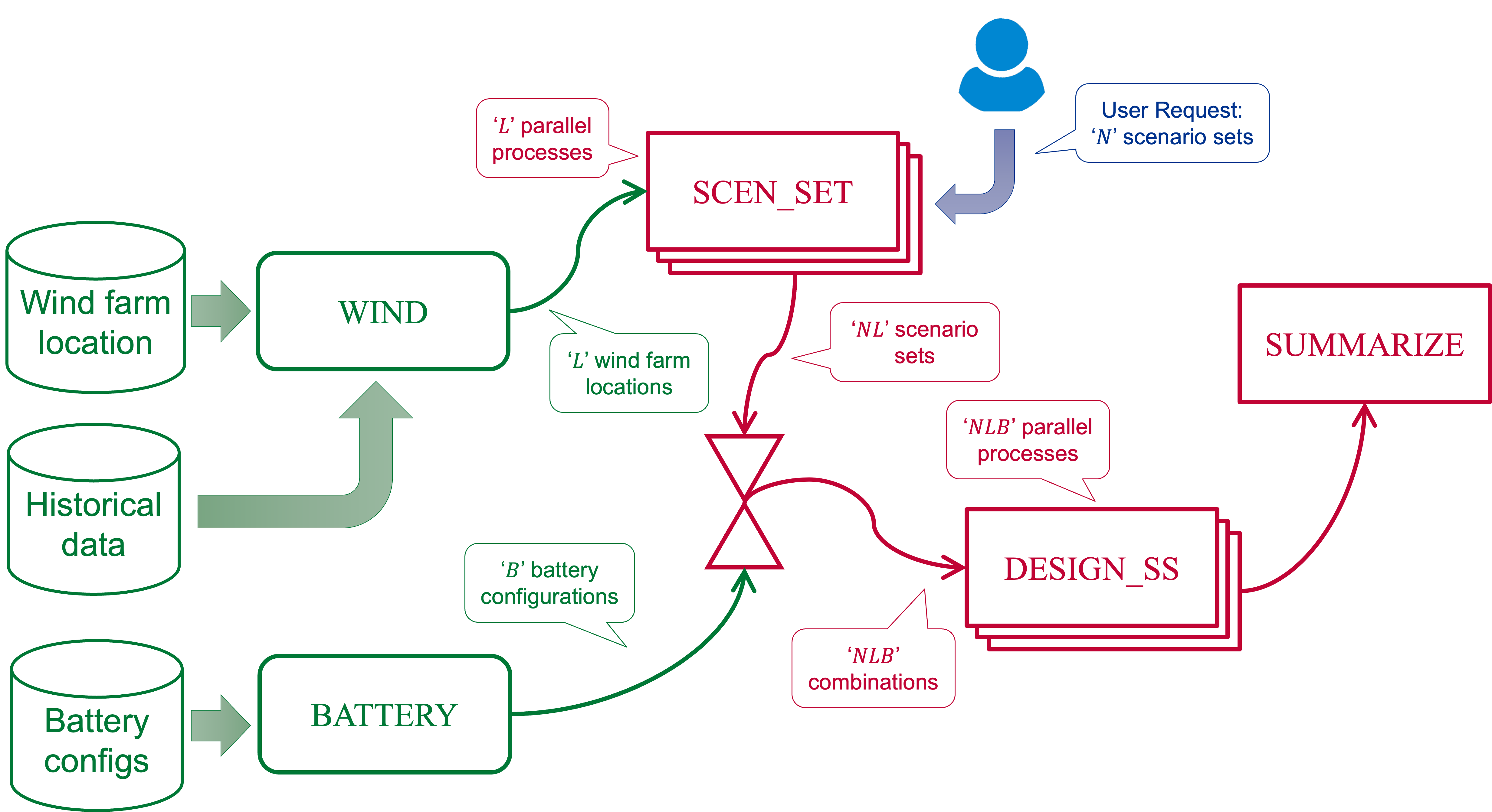}
    \caption{Summary of the \nextflow workflow for Formulation A constructed using the different processes.}
    \label{fig:workflow-v1}
\end{figure}

The first formulation focuses on creating multiple stochastic \emph{scenario sets} and evaluating an optimal solution for each each of them.
Each stochastic \emph{scenario set} comprises of randomly sampled historical wind speed and energy price data for $10$ representative days.
\cameo extracts the historical data set corresponding to each of the $5$ wind farm locations. 
Thereafter, it generates multiple \emph{scenario sets} (in this example, we limited our choice to $10$). 
Then, it generates combinations of battery parameters ($2$ chemistry, $4$ duration and $2$ ratings) with the $10$ scenario sets generated for each of the $5$ wind farm locations.
This results in $800$ input parameter combinations which are executed simultaneously using available compute resources.

The workflow for this formulation is summarized in Figure~\ref{fig:workflow-v1}.
The process \wind extracts the relevant wind farm location data and relevant historical information. 
This is sent to multiple parallel processes \sset to generate multiple random scenario sets for each location.
These scenario sets are combined with battery configurations extracted using the \battery process.
These combinations are sent to \dsset process to be executed in parallel.

\begin{figure*}[ht]
    \centering
    \includegraphics[width=0.33\textwidth]{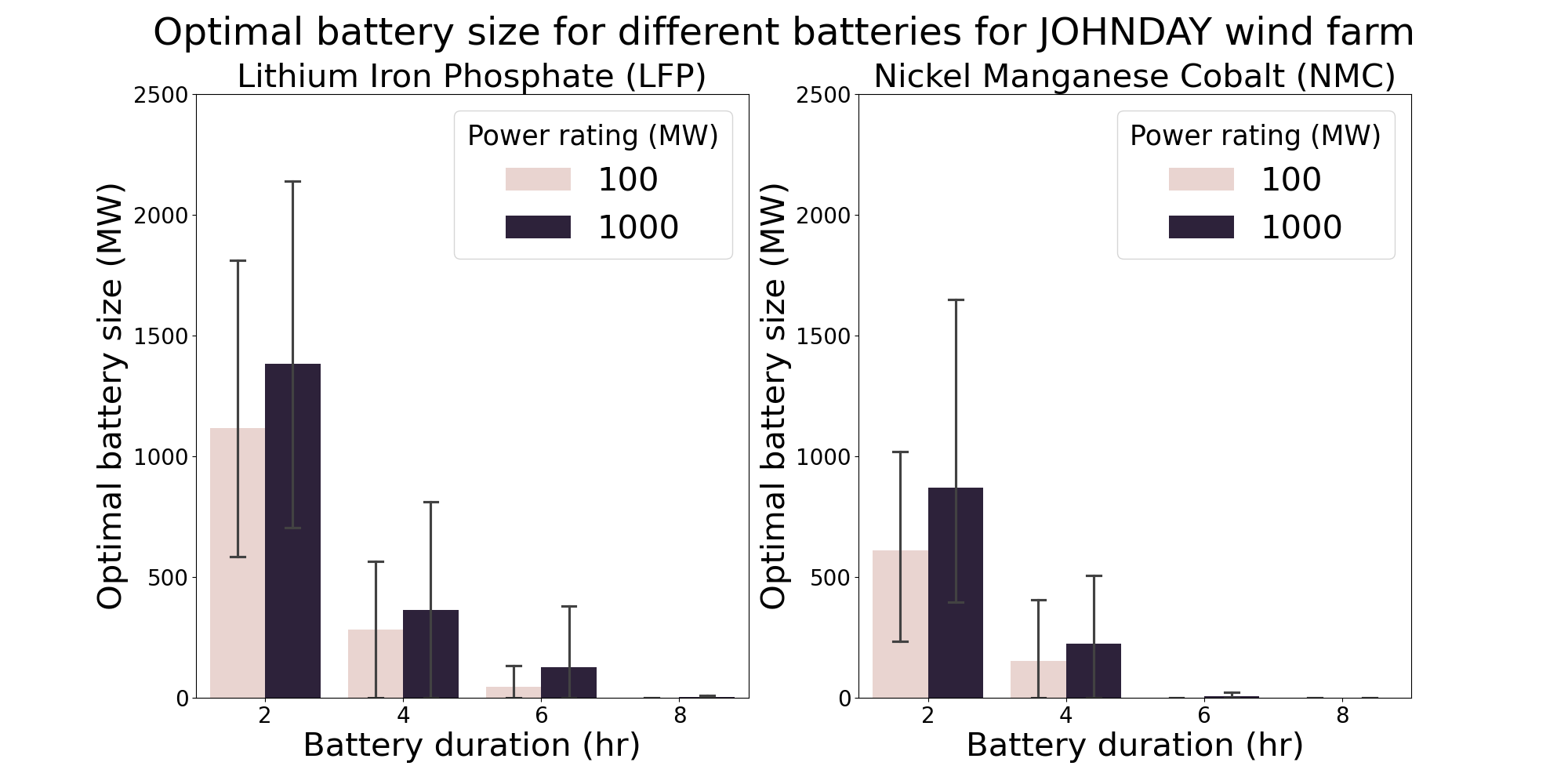}
    \includegraphics[width=0.33\textwidth]{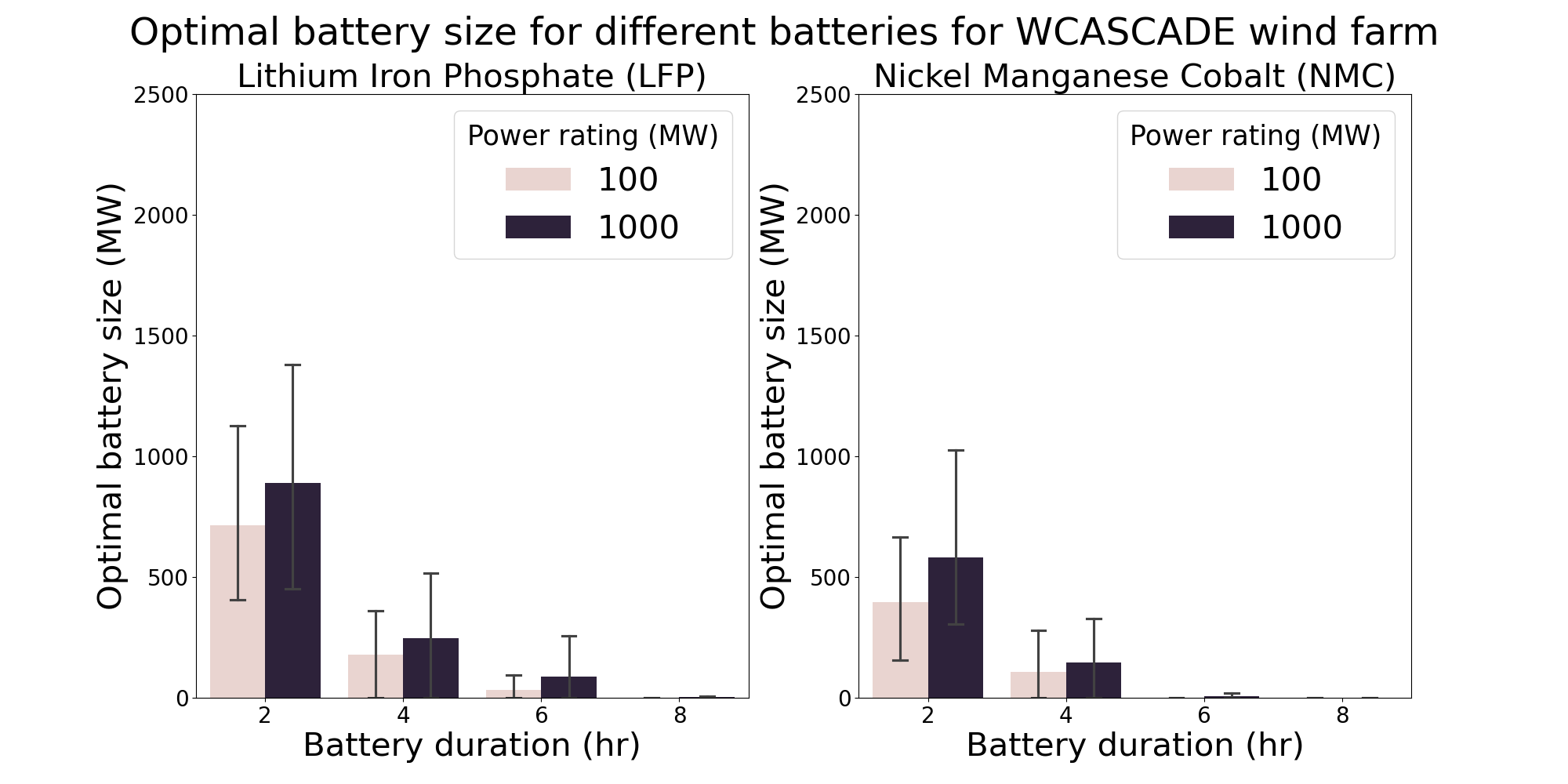}
    \includegraphics[width=0.33\textwidth]{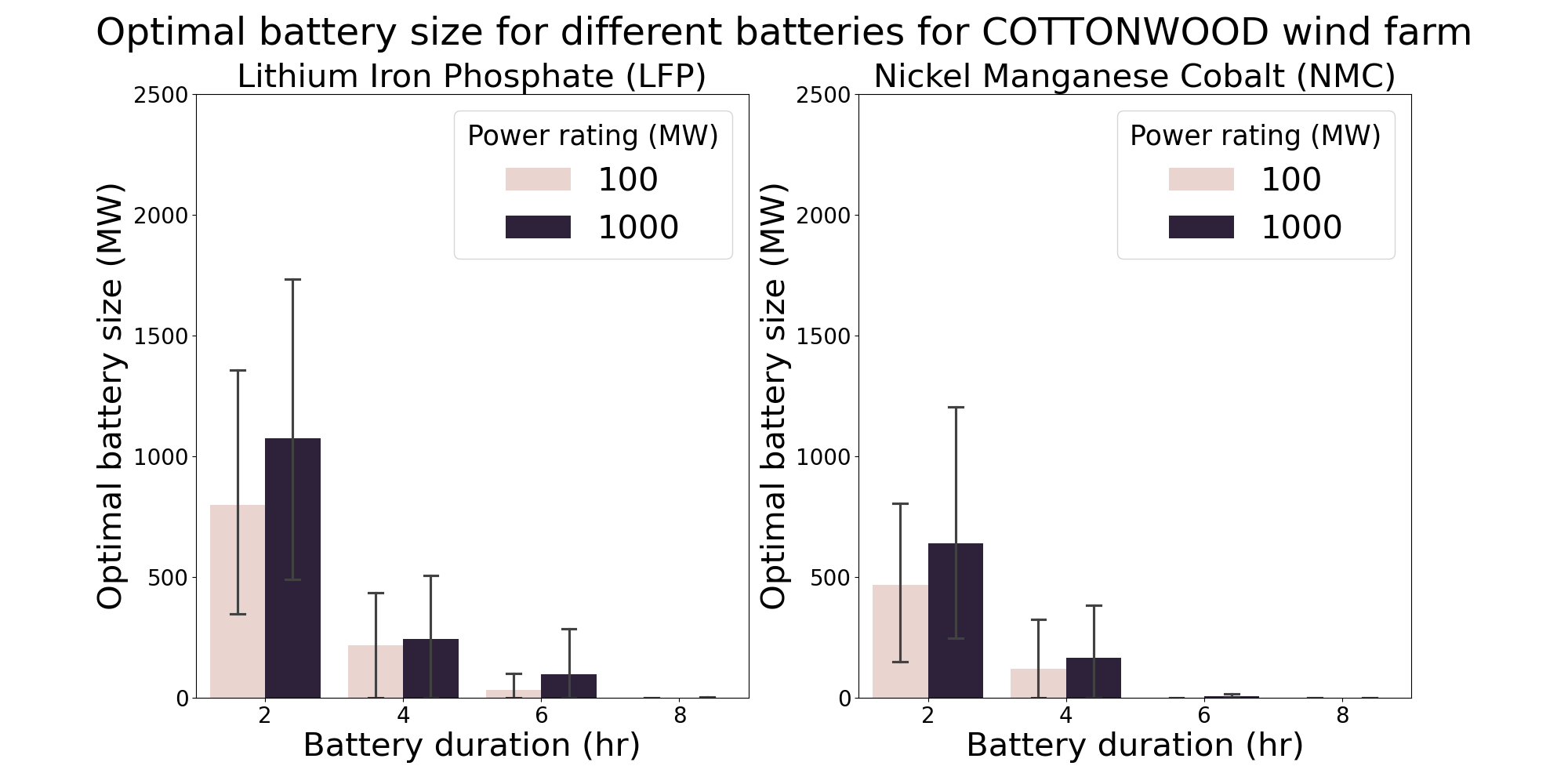}
    \includegraphics[width=0.33\textwidth]{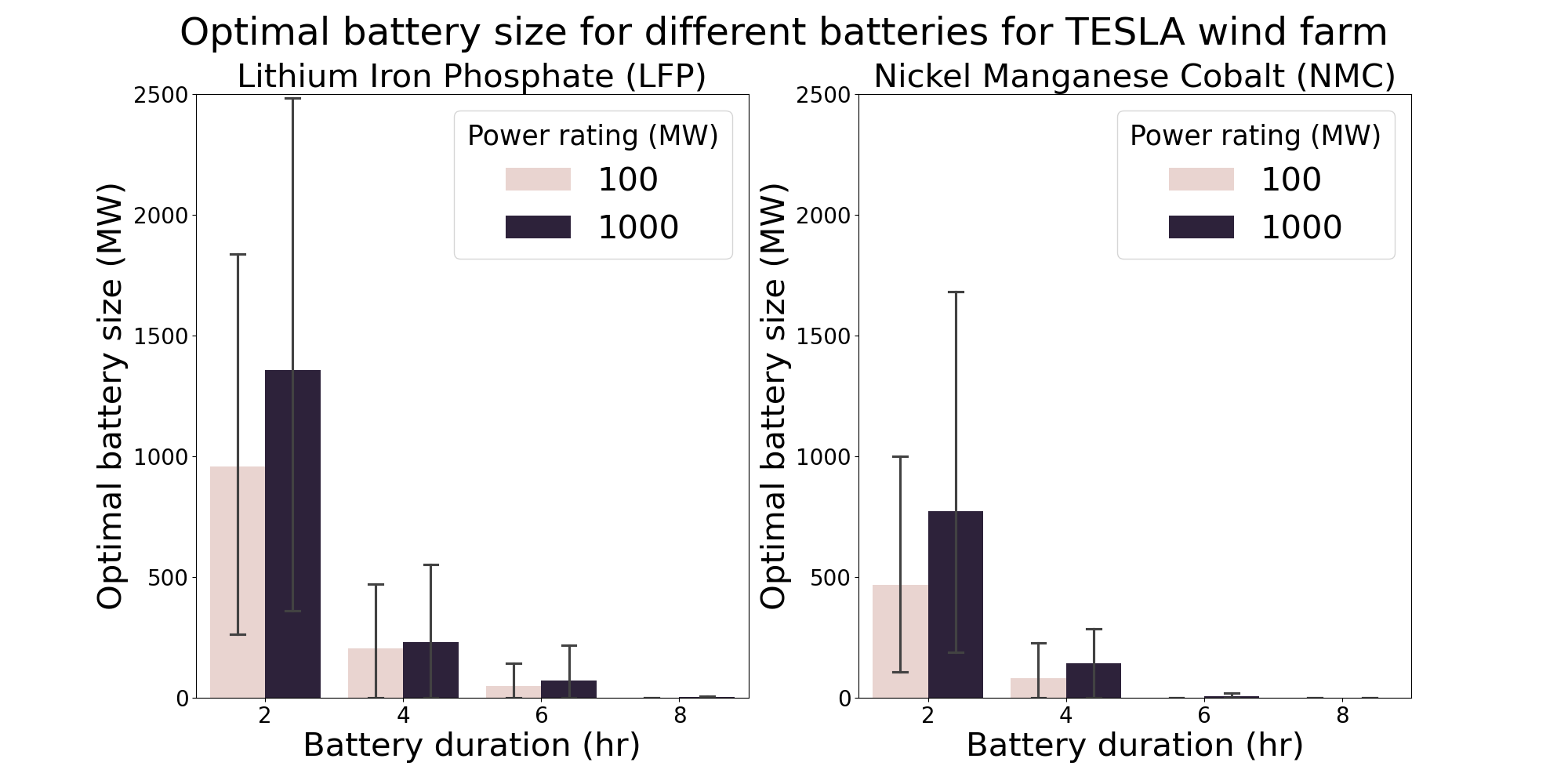}
    \includegraphics[width=0.33\textwidth]{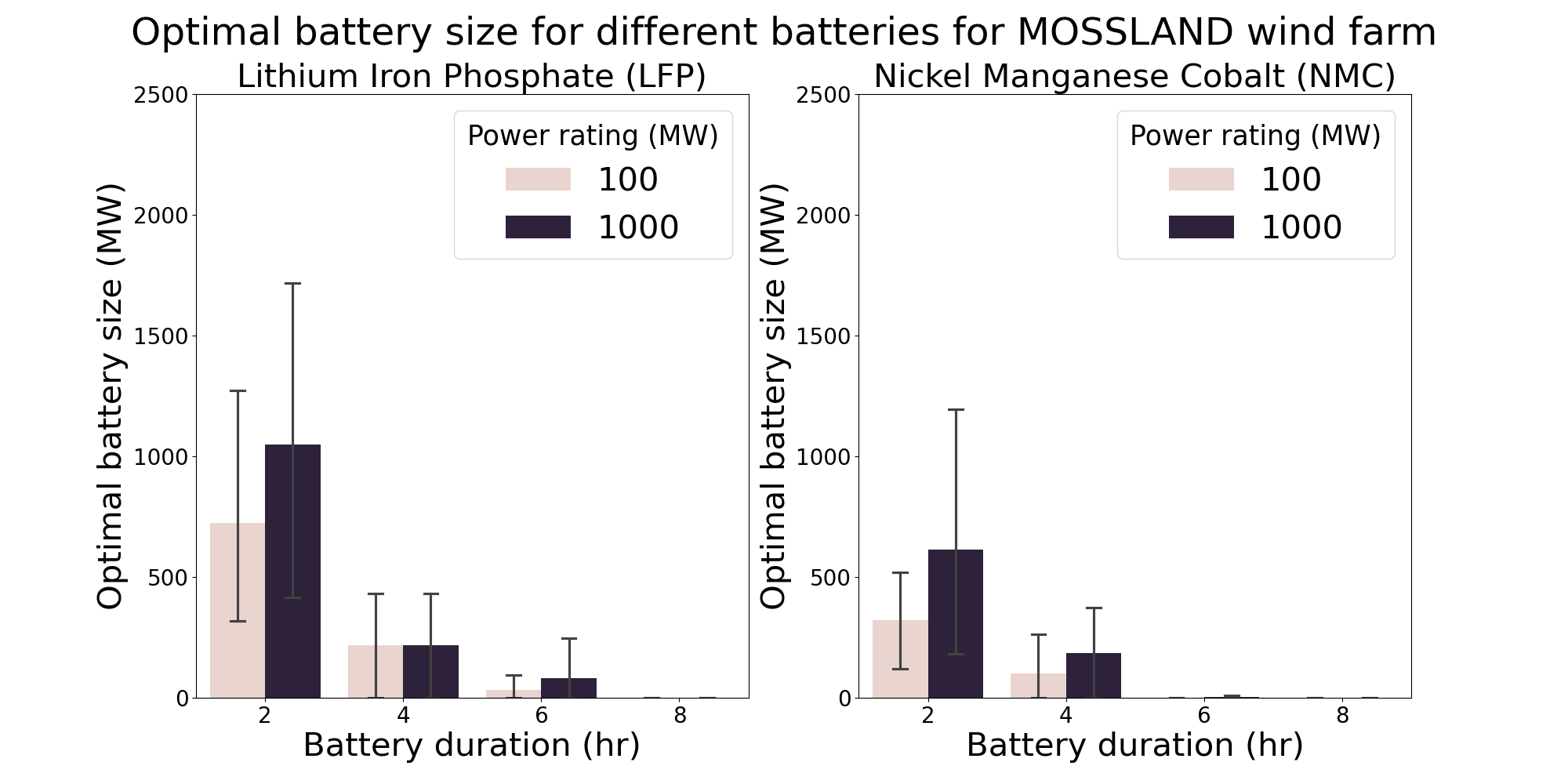}
    \caption{Optimal battery size for 5 different wind farm locations and 2 different battery chemistry. An optimization problem (based on Formulation A) is solved for different battery duration and ratings, which alters the initial installation investment of the batteries.}
    \label{fig:results-form1}
\end{figure*}
\begin{figure}[ht]
    \centering
    \includegraphics[width=0.48\textwidth]{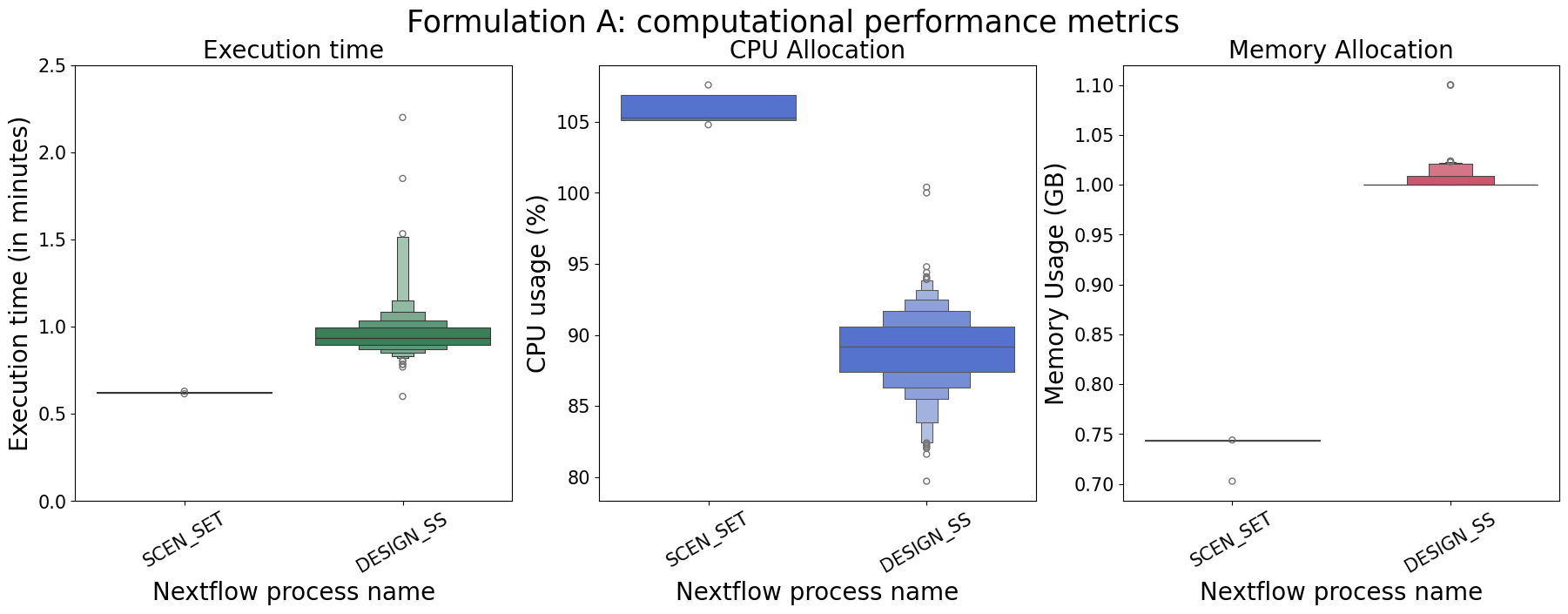}
    \caption{Statistics of performance metrics for Formulation A showing the job duration, CPU usage, and memory allocated for the different \nextflow processes in the workflow.}
    \label{fig:results-comp1}
\end{figure}
Finally, we summarize the findings using \summary and visualize the results through the plot shown in Figure~\ref{fig:results-form1}. 
Each of the $5$ panels represents a wind farm location and consists of $2$ plots for the different battery chemistry.
Each plot has $4$ sets of bars for the $4$ battery duration ($2,4,6$ and $8$ hours), with $2$ legends for the battery power ratings of $100$ and $1000$MW.
The error bar represents the variation in evaluation for the different stochastic scenario sets considered for each input parameter configuration.

We notice in Figure~\ref{fig:results-form1} that the optimal battery size evaluated for some battery configurations is negligible for all wind farm locations.
This is because of its high installation cost which cannot be significantly recovered by participating in the local energy market over the span of $30$ years. 

Additionally, we obtain the provenance report for all the executed processes at the end of the workflow.
Figure~\ref{fig:results-comp1} shows the performance metrics of the different processes used in the workflow.
These metrics include the execution time (in minutes), the \% CPU allocation and the memory allocation (in GB).
Note that the metrics are shown separately for each process. 
In Formulation A, \cameo executed $5$ \sset processes and $800$ \dsset processes. We notice that the average execution time of the design optimization process \dsset is around $1$ minute.

\subsection{Results from Formulation B.}
\begin{figure}[tbhp]
    \centering
    \includegraphics[width=0.48\textwidth]{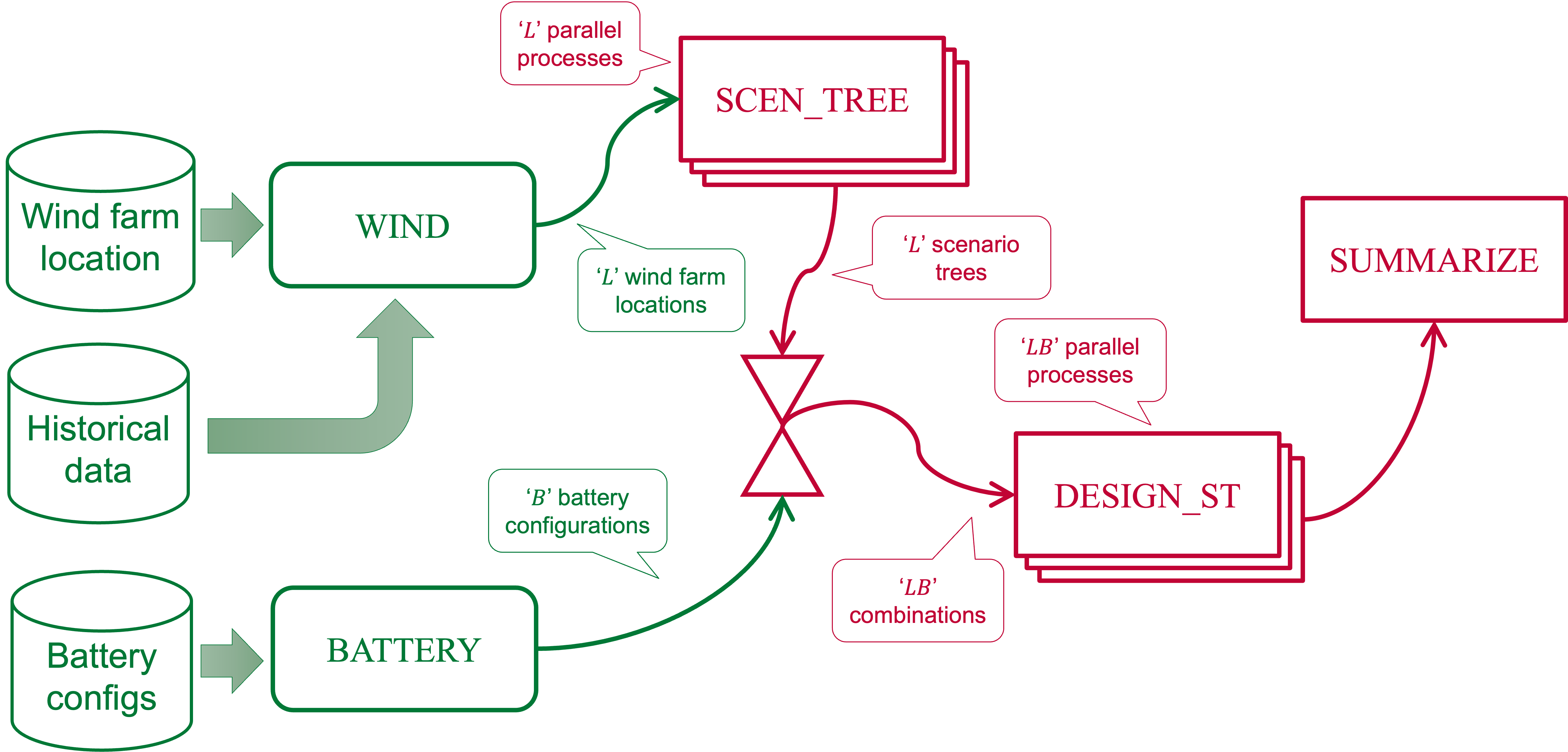}
    \caption{Summary of the \nextflow workflow for Formulation B constructed using the different processes.}
    \label{fig:workflow-v2}
\end{figure}
This formulation considers a multi-stage optimization problem for each wind farm location.
The workflow for this formulation is shown in Figure~\ref{fig:workflow-v2}.
To this end, the \wind process extracts relevant information for each of the $5$ wind farm locations.
Subsequently, the extracted data is sent to the \stree process, which generates a \emph{scenario tree} corresponding to each wind farm location.
These $5$ \emph{scenario trees} are combined with the $16$ battery configurations considered for the earlier formulation ($2$ battery chemistry, $4$ battery duration and $2$ battery power ratings).
This leads to a total of $80$ optimization problems that are executed in parallel.
The \dstree process performs this \emph{scenario tree} multi-stage optimization task.
\begin{figure*}[tbhp]
    \centering
    \includegraphics[width=0.33\textwidth]{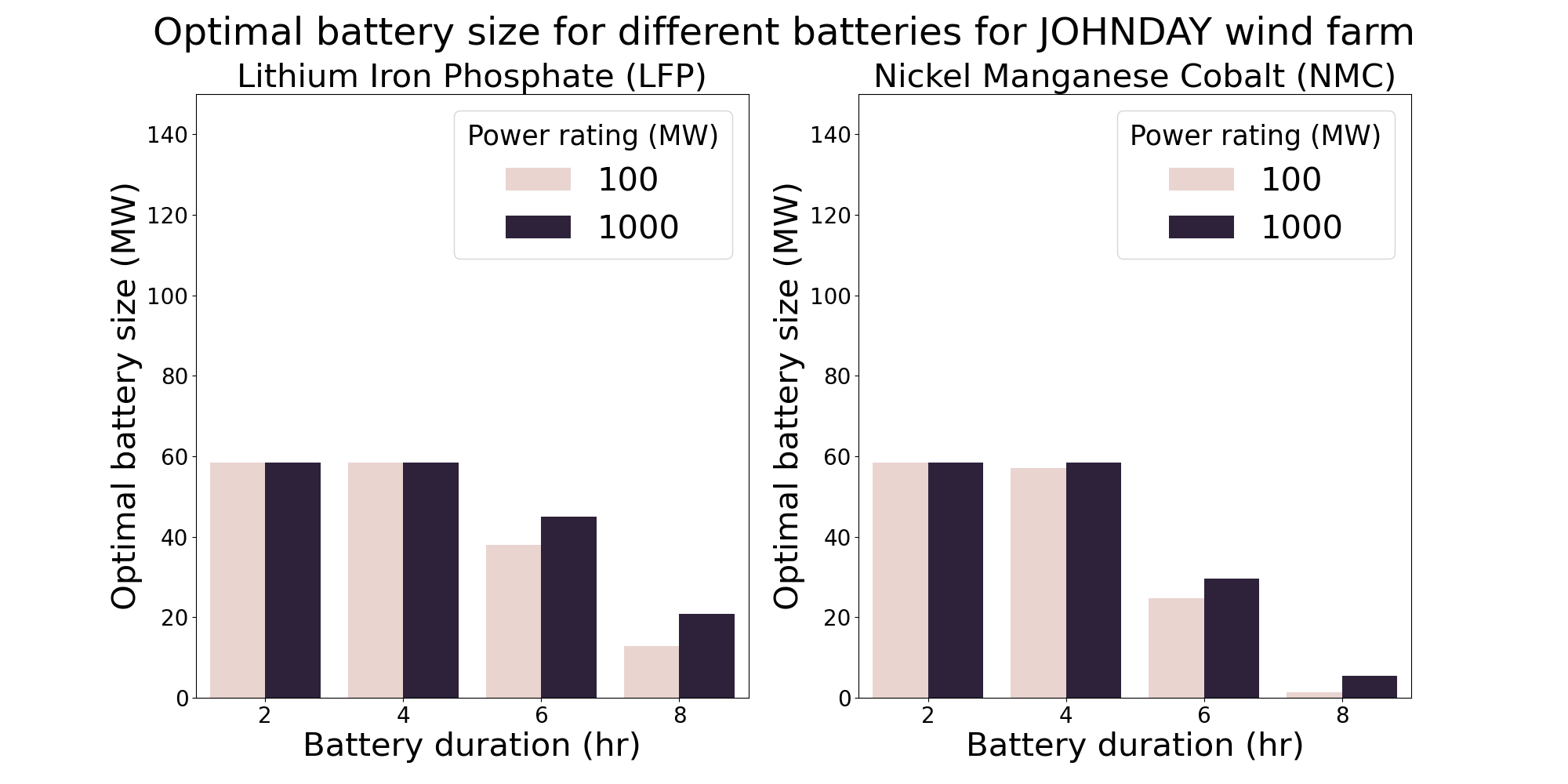}
    \includegraphics[width=0.33\textwidth]{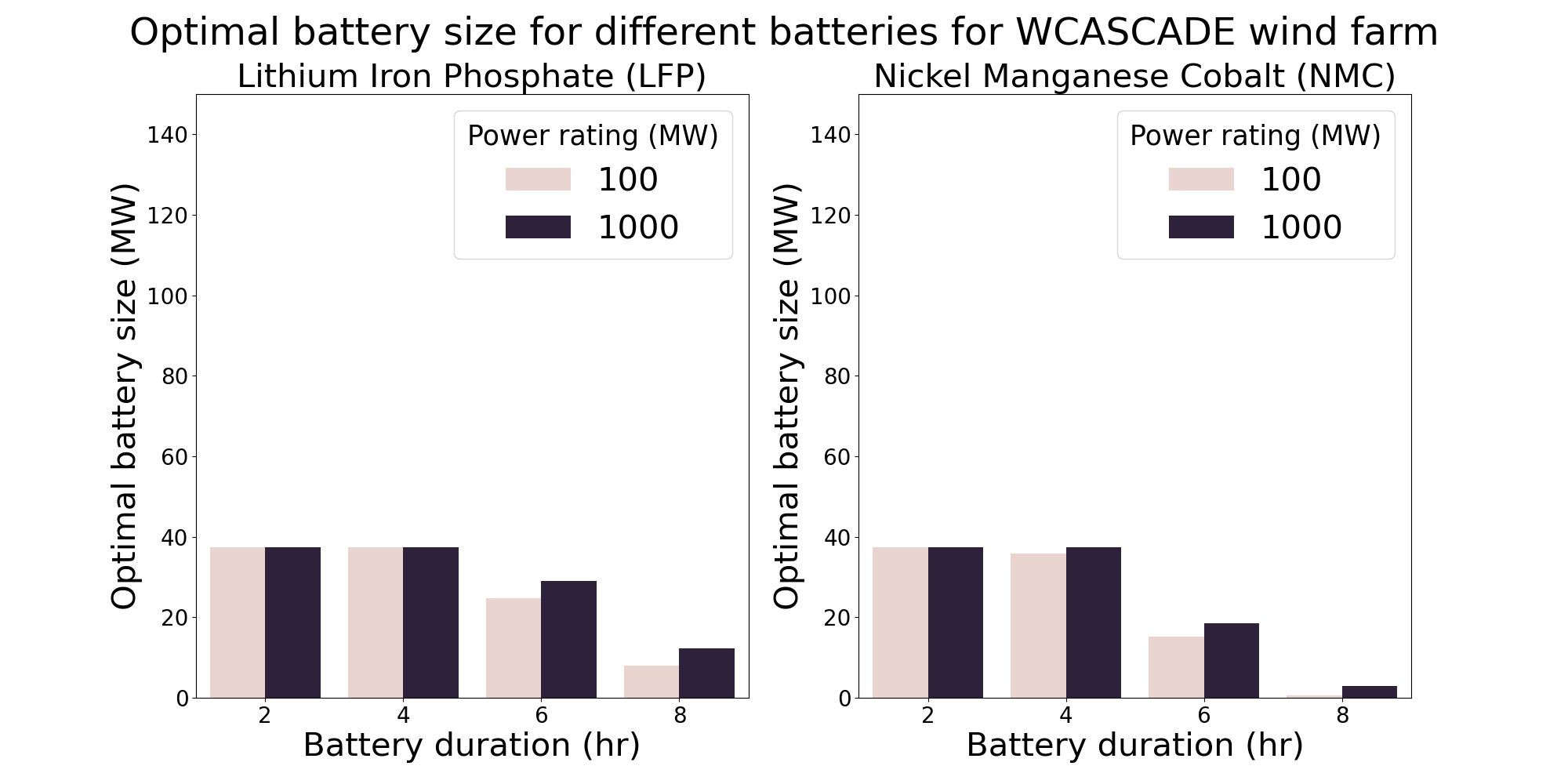}
    \includegraphics[width=0.33\textwidth]{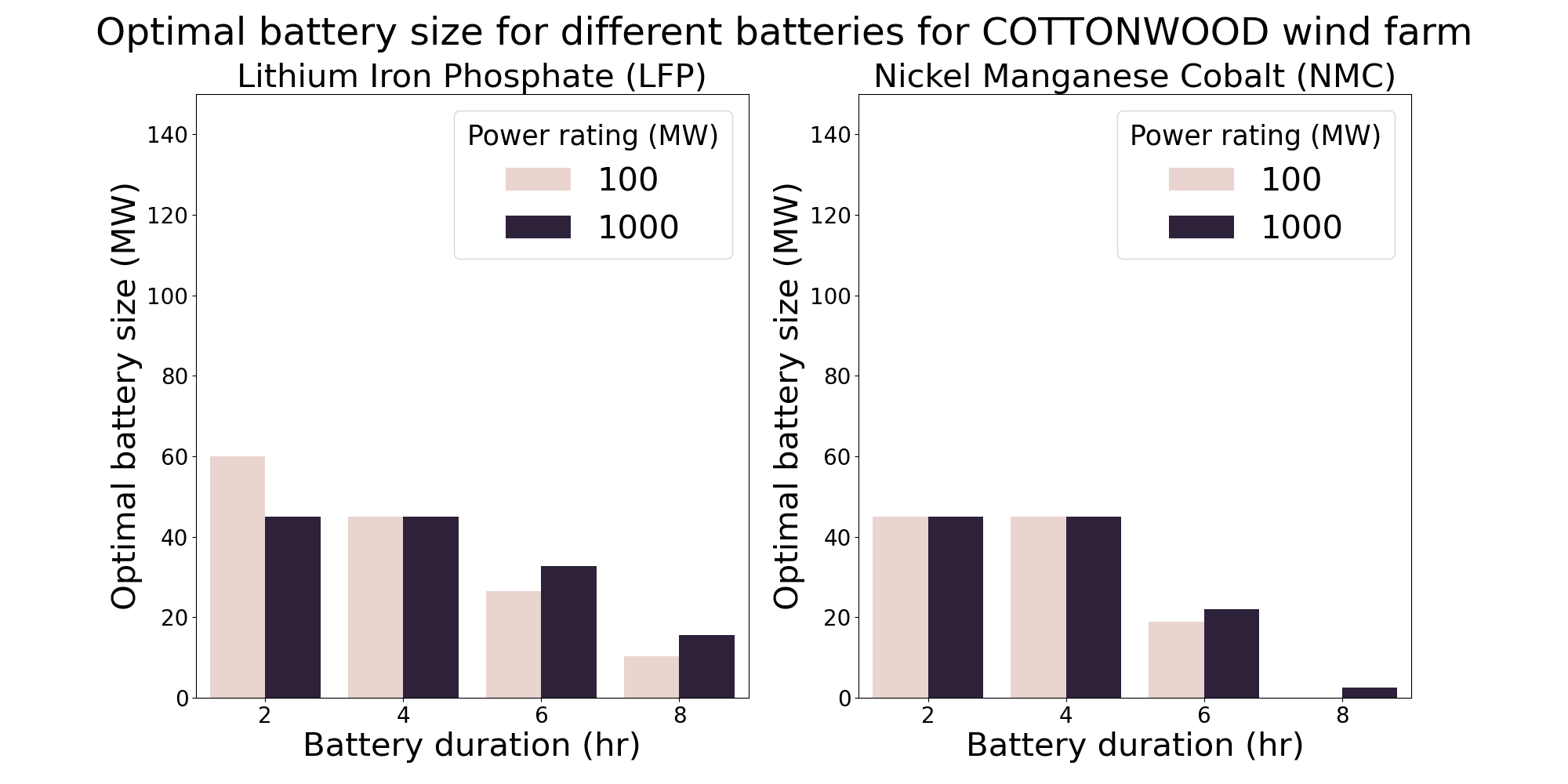}
    \includegraphics[width=0.33\textwidth]{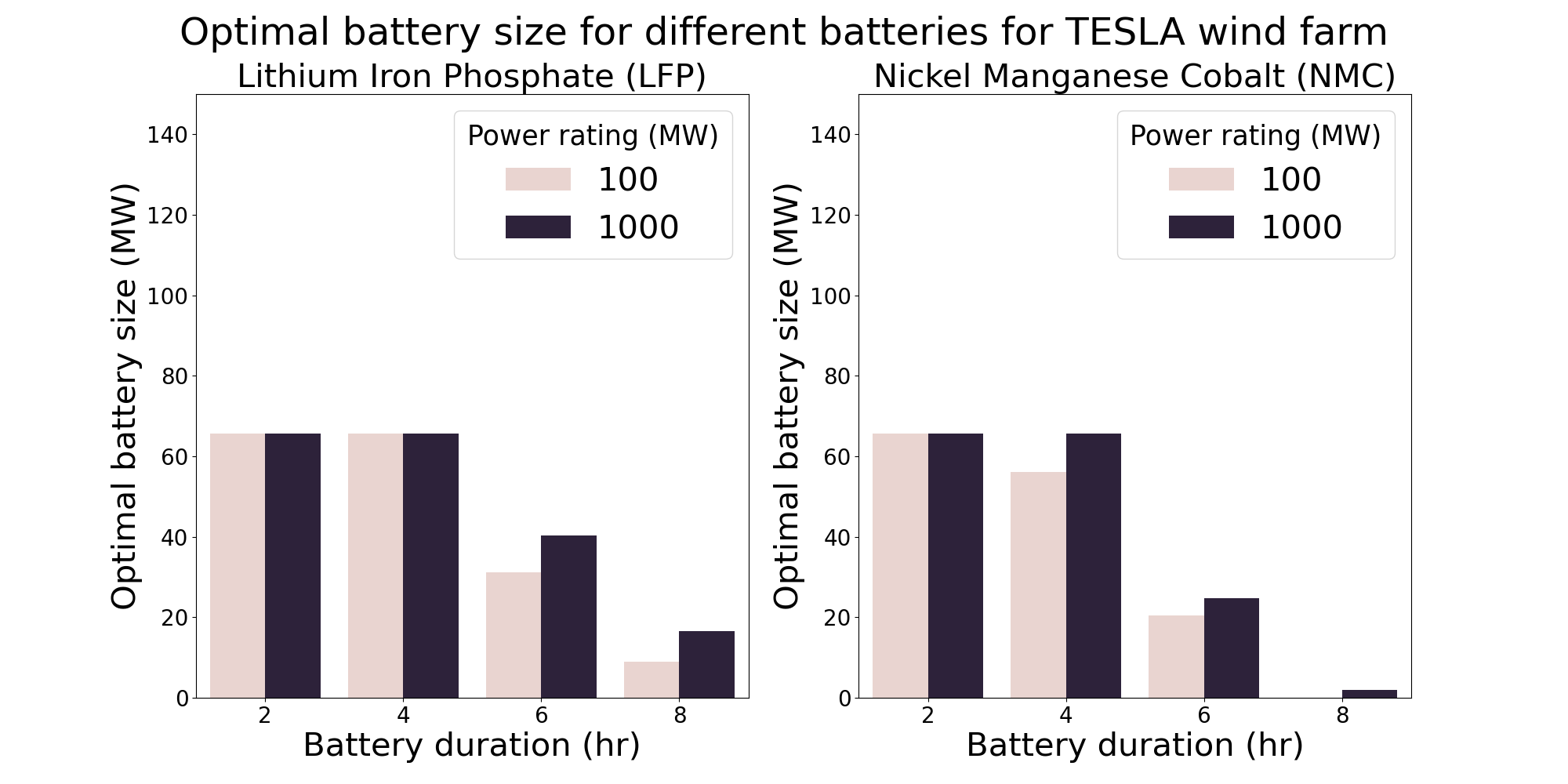}
    \includegraphics[width=0.33\textwidth]{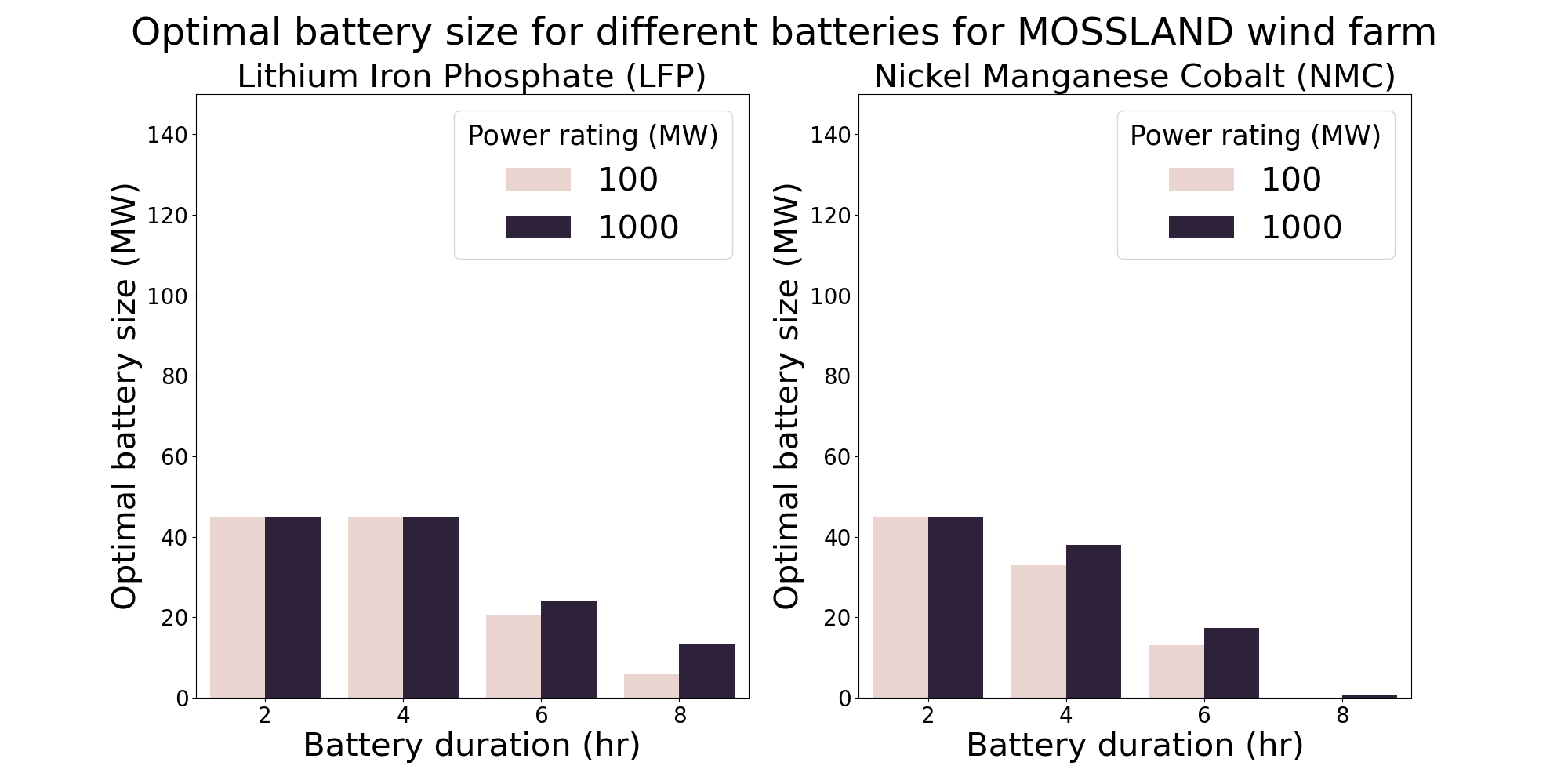}
    \caption{Optimal battery size for 5 different wind farm locations and 2 different battery chemistry. A multi-stage optimization problem (based on Formulation B) is solved for different battery duration and ratings, which alters the initial installation investment of the batteries.}
    \label{fig:results-form2}
\end{figure*}

Finally, the results are summarized in a consolidated CSV file using the \summary process and relevant plots are visualized as shown in Figure~\ref{fig:results-form2}.
Note that the error bars in the bar plot are missing. This is because unlike Formulation A, here we do not consider multiple random scenario sets.
Rather, we obtain a single \emph{scenario tree} for each wind farm location and solve a multi-stage optimization problem using it. 

\begin{figure}[tbhp]
    \centering
    \includegraphics[width=0.48\textwidth]{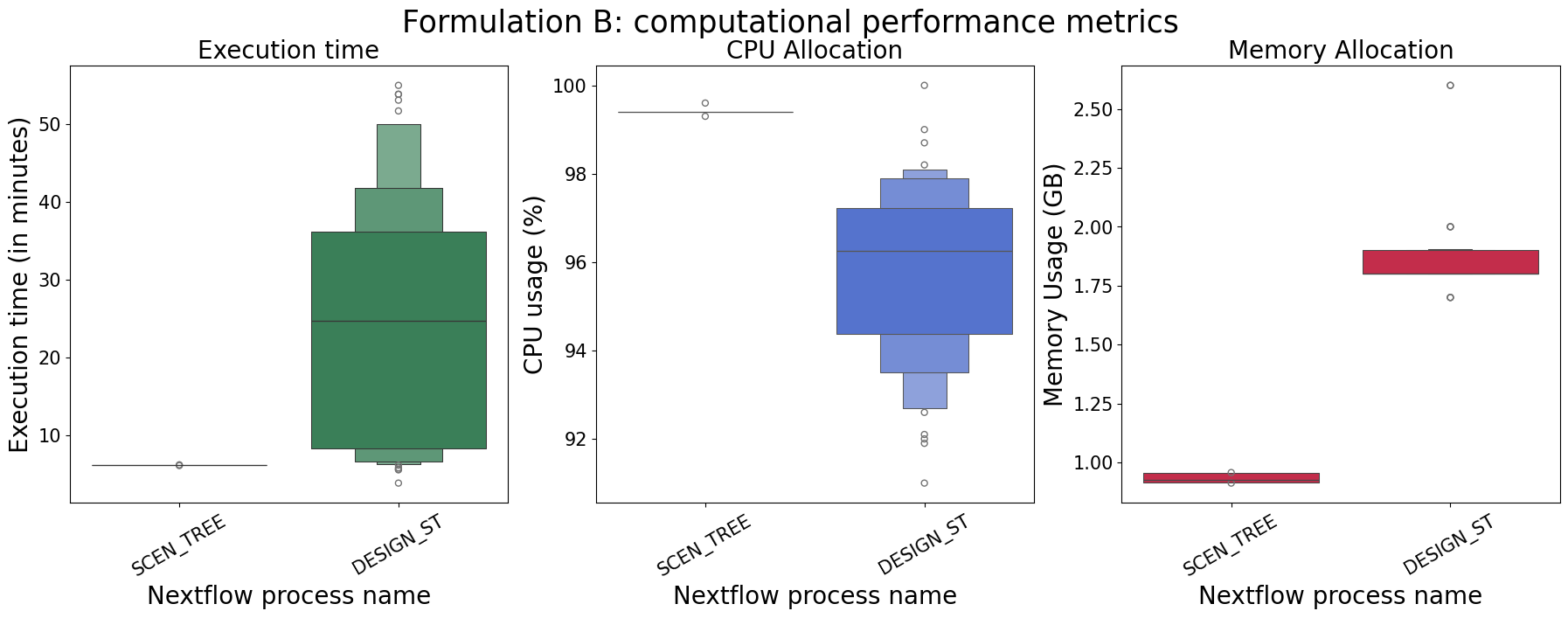}
    \caption{Statistics of performance metrics for Formulation B showing the CPU usage, memory allocated and job duration for the different \nextflow processes in the workflow.}
    \label{fig:results-comp2}
\end{figure}
The computational performance metrics are also displayed as a provenance report as shown in Fig.~\ref{fig:results-comp2}.
In this case, \cameo executes $5$ instances of \stree process and $80$ instances of $\dstree$ process.
We note that unlike Formulation A, both the processes (\stree and \dstree) consume significantly more computation time as well as use larger memory resources.
This analysis enables users to characterize the computational overhead required for each formulation, so that required resources can be allocated for future implementations.

\section{Conclusion}
As more renewable are added to the energy system, it creates challenges for designing and operating a reliable, resilient power system. Co-design of such a system requires multi-disciplinary technical capabilities from system modeling, optimization, and high performance computing. This paper presents a novel architecture to facilitate multi-objective co-design, leveraging modular workflow systems and high performance computing resources. We also present a power electronics use case to design an offshore wind farm with economic and operational objectives.

In the future, we plan to implement more features in the \cameo architecture to provide available library of modules and processes for common co-design problems, add a graphic user interface for easier user interaction, and enable interfacing with multiple execution environments.

\section*{Acknowledgement}
The research was supported by the Energy System Co-Design with Multiple Objectives and Power Electronics (E-COMP) Initiative at Pacific Northwest National Laboratory (PNNL). It was conducted under the Laboratory Directed Research and Development Program at PNNL.

\printbibliography

\end{document}